%7/97
\magnification=1200
\baselineskip=18pt

\centerline{\bf LORENTZ-INVARIANT HAMILTONIAN AND}
\centerline{\bf  
RIEMANN HYPOTHESIS}

\bigskip

\centerline{by}

\bigskip

\centerline{Susumu Okubo}
\centerline{Department of Physics and Astronomy}
\centerline{University of Rochester}
\centerline{Rochester, NY 14627, U.S.A.}

\vskip 2truein

\noindent \underbar{\bf Abstract}

We have given some arguments that
 a two-dimensional Lorentz-invariant Hamiltonian  
may be relevant to the Riemann hypothesis concerning zero points of the 
Riemann zeta function. Some eigenfunction of the Hamiltonian corresponding 
to infinite-dimensional representation of the Lorentz group have many 
interesting properties. Especially, a relationship exists between the zero 
zeta function condition and the absence of trivial representations in the 
wave function.

\vfill\eject

The Riemann hypothesis ([1], [2], [3]) is one of the long-standing 
problems in the number theory. The Riemann's zeta function $\zeta (z)$ for 
a complex variable $z$ is defined for Re $z > 1$ by
$$\zeta (z) = \sum^\infty_{n=1} {1 \over n^z}$$
and for other values of $z$ by its analytic continuation. It is well-known 
that $\zeta (z)$ is zero for negative even integer values of $z$, i.e.
 $z = -2,-4,-6,\dots$, while all other non-trivial zeros of $\zeta 
(z)$ must lie in the strip $0 < \ {\rm Re}\ z < 1$.  It has been 
conjectured that all non-trivial zeros of $\zeta (z)$ actually lie on the 
critial line Re $z = {1 \over 2}$. This Riemann hypothesis (hereafter 
referred to as RH) is important in the number theory, since its validity 
can answer some questions concerning distributions of the prime numbers.

It has been suggested by many authors that the problem may be related to 
eigenvalue spectra of a self-adjoint operator H in some Hilbert space, 
although any such H has not been found so far. This view has been 
strengthened by the works of Odlyzko [4] and of others (see e.g. [5] and [6
], and references quoted therein) that the statistical distributions of 
zero points of $\zeta (z)$ is consistent to a high degree with the law of 
the Gaussian unitary ensemble of random matrix theory [5], which is 
expected for spectra of complex Hamiltonians. Moreover, this fact is
also  found 
 to be related to the phenomenon of the quantum chaos ([6] and [7]). A 
widely held opinion among many authors is that the validity of RH with its 
assoiciated Hamiltonian, if it exists, will shed light to the quantum 
chaos and vice versa.

The purpose of this note is to show the existence of one-parameter family 
of complex Hamiltonians which seems to be intimately connected with the 
problem. Moreover, these Hamiltonians are invariant under two-dimensional 
Lorentz transformation, a fact which will be of some intrinsic interest 
for its own right.

We start from the following integral representation [8] of $\zeta (z)$:
$$\zeta (z) = {1 \over (1-2^{1-z})\Gamma (z)} \int^\infty_0 dt {t^{z-1} 
\over 1 + \exp t} \quad , \quad ({\rm Re}\ z > 0) \eqno (1)$$
so that any non-trivial zero of $\zeta (z)$ must satisfy
$$\int^\infty_0 dt {t^{z-1} \over 1 + \exp t} = 0 \quad , \quad (1 > 
\ {\rm Re}\ z > 0) \quad . \eqno(2)$$
We call the condition Eq. (2) be zero zeta function condition (hereafter 
abbreviated as ZZFC).  Also in view of the identity
$$2^{1-z} \Gamma (z) \zeta (z) \cos \big( {\pi \over 2} z \big) = \pi^z \zeta
(1-z) \quad , \eqno(3)$$
we may restrict ourselves to consideration only of the half sector $1 >\ {
\rm Re}\ z \geq {1 \over 2}$ instead of $1 > \ {\rm Re}\ z > 0$ for ZZFC.
Especially, if we can show that the assumption of $1 > \ {\rm Re}\ z > {1
 \over 2}$ for $\zeta (z) =0$ will lead to a contradiction, this will prove 
RH.

Suppose now that a Hamiltonian H is hermitian in a Hilbert space, so that 
we have
$$<H \phi |\psi>\  =\  < \phi |H \psi> \eqno(4)$$
for wave functions $\phi$ and $\psi$. For a complex $z$ satisfying ZZFC, 
i.e. Eq. (2), we set 
$$z = {1 \over 2} + i \lambda \quad . \eqno(5)$$
If H possessses a eigenfunction $\phi_0$ with the eigenvalue $\lambda$, 
i.e. if we have
$$H \phi_0 = \lambda \phi_0 \quad , \eqno(6)$$
then Eq. (4) with $\psi = \phi = \phi_0$ will give $\lambda = \overline 
\lambda$ being real and hence Re $z = {1 \over 2}$, proving RH. The natural 
question is whether such a H exists or not. Although we could not 
completely succeed, we have found some pairs $(H, \phi_0)$ satisfying the 
required condition Eq. (6), almost proving RH. The problem is that $\phi_0$ 
found so far appear to be not normalizable.  However, a possibility exists 
that H may possess the correct eigenfunction  $\phi_0$. Moreover, there 
exists an intriguing connection between ZZFC and the representation space of 
the Lorentz group under which H is invariant.  These facts suggest that 
our Hamiltonian H may indeed  still be  relevant to the problem of RH.

Let $\phi (x,y)$ and $\psi (x,y)$ be functions of two real variables $x$ 
and $y$. We introduce the inner product by
$$< \phi |\psi>\  = \int^\infty_{- \infty} dx \int^\infty_0 dy\  \overline \phi 
(x,y) \psi (x,y) \quad . \eqno(7)$$
Here and hereafter, $\overline \phi (x,y)$, for example, stands for the 
complex conjugate of $\phi (x,y)$. Note that the ranges of the integrations 
are $\infty > x > - \infty$ for $x$ but $\infty > y \geq 0$ for $y$. 
Consider a family of second-order differential operators given by
$$H = {\partial^2 \over \partial x \partial y} + i \beta y {\partial \over 
\partial y} + i (1 - \beta) x {\partial \over \partial x} + {i \over 2} \eqno
(8)$$
for real parameter $\beta$. We note first that H is complex rather than 
real and second that it contains a purely imaginary constant term $i/2$ 
whose presence is crucial for the hermiticity property of H, as we will see 
below. By a simple calculation, it is easy to find
$$\overline{(H \phi)} \psi - \overline \phi (H \psi ) = {\partial \over 
\partial x} J_1 + {\partial \over \partial y} J_2 \eqno(9a)$$
where we have set
$$\eqalignno{J_1 &= {1 \over 2} \bigg( {\partial \overline \phi \over
\partial y} \psi - \overline \phi {\partial \psi \over \partial y}
\bigg) - i (1 - \beta) x \overline \phi \psi &(9b)\cr
J_2 &= {1 \over 2} \bigg( {\partial \overline \phi \over
\partial x} \psi - \overline \phi {\partial \psi \over \partial x}
\bigg) - i \beta y \overline \phi \psi \quad . &(9c)\cr}$$
Note that the presence of the constant term $i/2$ in the right side of Eq. 
(8) is pivotal in enabling to obtain Eqs. (9). Integrating both sides of 
Eq. (9), we will find the hermiticity condition Eq. (4), if we could discard 
all partially integrated terms involving $J_1$ and $J_2$.  From the 
explicit expressions of $J_1$ and $J_2$ given above, this would be 
possible, if $\phi$ and $\psi$ or their derivatives with respect to $x$ 
vanish at $y=0$, and if $\phi$ and $\psi$ as well as their derivatives 
decrease sufficiently rapidly for $x \rightarrow \pm \infty$ and $y 
\rightarrow \infty$. Of course, we have to more carefully study the question 
of the domain and range of H in order to establish the self-adjointness of 
H. However, the naive criteria given above suffices for the present 
discussion. Especially, if $\phi$ satisfies Eq. (6), i.e.
$$H \phi = \lambda \phi \eqno(10a)$$
with the boundary condition
$$\phi (x,0) = 0 \eqno(10b)$$
at $y=0$  and if $\phi (x,y)$ decreases rapidly at infinity, we will be 
able to establish RH in principle. We note that Eq. (10a) with $z = {1 
\over 2} + i \lambda$ implies the validity of
$$\bigg\{ {\partial^2 \over \partial x \partial y} + i \beta y {\partial \over 
\partial y} + i (1 - \beta) x {\partial \over \partial x} \bigg\} \phi = - i z 
\phi \quad . \eqno(11)$$
We have yet to meaningfully utilize ZZFC in our formalism. Before going 
into its detail, we will first, however, note the following property of the 
Hamiltonian. H as well as the inner product $<\phi |\psi>$ are clearly 
invariant under the transformation
$$x \rightarrow {1 \over k} x \quad , \quad {\rm and} \quad y \rightarrow ky 
\eqno(12)$$
for any positive constant $k$. This invariance really reflects  that of
 two-dimensional Lorentz transformation. To understand it better, consider new 
variables $u$ and $v$ given by
$$x = u - v \quad , \quad y = u + v \quad . \eqno(13)$$
The Hamiltonian H is then invariant under the SO(1,1) Lorentz 
transformation
$$\eqalign{u \rightarrow u^\prime &= (\cosh \theta) u + (\sinh \theta) v 
\quad ,\cr
v \rightarrow v^\prime &= (\sinh \theta) u + (\cosh \theta) v
\cr}\eqno(14)$$
for real constant $\theta$, corresponding to the boost parameter $k = \exp
\theta$. Because of the invariance, if $\phi$ satisfies $H \phi = \lambda 
\phi$, then so does $\phi ({x \over k}, ky)$, and hence
$$\widetilde \phi (x,y) = \int^\infty_0 {dk \over k} f (k) \phi \big({x \over k}, 
ky\big) \eqno(15)$$
for arbitrary function $f(k)$ satisfies also $H \widetilde \phi = \lambda
 \widetilde \phi$.  Especially, any eigenfunction $\phi (x,y)$ of H may be 
regarded as a infinite-dimensional realization of the Lorentz group
 SO(1,1).

After these preparations, we will now discuss solutions of the differential 
equation (11). We have found the following two families of solutions. Let 
$g(\xi)$ be an arbitrary function of a variable $\xi$ which vanishes fast 
for $\xi \rightarrow \infty$. Then, we show first that
$$\phi (x,y) = \int^\infty_0 dt\  t^{z-1} \exp \{ ix t^{1-\beta}\} g(t + y
t^\beta)\quad , \quad ({\rm Re}\ z>0) \eqno(16)$$
with $\xi = t + y t^\beta$ is a solution of $H \phi = \lambda \phi$ 
with $z = {1 \over 2} + i \lambda$.  In this connection if we change $x 
\leftrightarrow y$ and $\beta \leftrightarrow 1 - \beta$, it will also 
furnish a solution.  This can be proved as follows. For simplicity, set
$$G_0 (x,y;t) = \exp \{ ix t^{1-\beta}\} g(t+y t^\beta) \quad \eqno(17a)$$
and note that $G_0$ satisfies a differential equation
$$\bigg\{ {\partial^2 \over \partial x \partial y} + i \beta y {\partial \over 
\partial y} + i (1 -\beta) x {\partial \over \partial x} \bigg\} G_0 = i t {
\partial \over \partial t} G_0 \eqno(17b)$$
as we can easily verify.  Multiplying $t^{z-1}$ and integrating over $t$ 
from $t = \infty$ to $t=0$, then it reproduces Eq. (11)
 if Re $z>0$. Especially, the special choice of
$$g (\xi) = {1 \over 1 + \exp \xi}$$
is of interest. Then, the function $f_0$ given by 
$$f_0 (x,y) = \int^\infty_0 dt\  {t^{z-1} \over 1 + \exp [t+ y t^\beta]} \exp 
(ix t^{1 - \beta}) \eqno(18a)$$
obeys
$$H f_0 = \lambda f_0\quad , \eqno(18b)$$
although ZZFC implies only
$$f_0 (0,0) = 0 \eqno(19)$$
at the single point $x = y = 0$, but not the desired boundary condition Eq. 
(10b) for arbitrary $x$. As we will see later, $f_0 (x,y)$ is intimately
 related to the zeta function.

We can also find another class of solutions as follows. Let us consider now
$$G_1 (x,y;u) = {u^{\theta -1} \over (1-u)^\theta} e^{-iuxy} g(y u^\beta (1
-u)^{1-\beta}) \eqno(20)$$
for a constant $\theta$ with $\xi = y u^\beta (1-u)^{1 -\beta}$ for 
arbitrary function $g(\xi)$. We can verify that $G_1$ satisfies the 
differential equation
$$\bigg\{ {\partial^2 \over \partial x\partial y} + i \beta y {\partial \over
 \partial y} + i (1-\beta) x {\partial \over \partial x}\bigg\} G_1 - i {\partial
\over \partial u}\{ u(1-u)G_1\} = - i \theta G_1 \quad . \eqno(21)$$
Integrating Eq. (21) from $u=1$ to $u=0$, and assuming $1  > \ {\rm Re}\ 
\theta > 0$, it gives
$$H f_1 = \lambda_1 f_1 \quad , \eqno(22a)$$
with
$$\theta = {1 \over 2} + i \lambda_1 \quad , \eqno(22b)$$
if we set
$$f_1 (x,y) = \int^1_0 du\  G_1 (x,y,u) \quad . \eqno(22c)$$
In order to obtain a solution which satisfies Eq. (10b), we let 
$x \rightarrow {1 \over k(t)} x$ and $y \rightarrow k(t)y$ for an arbitrary 
function $k(t)$ of a new variable $t$, and integrate Eq. (22c) on $t$ 
after multiplying $t^{z-1} (1+ \exp t)^{-1}$. In this way, we generate a 
new family of solutions. In summary, the function
$$\phi_1 (x,y) = \int^\infty_0 dt\  {t^{z-1} \over 1 + \exp t} \int^1_0 du 
{u^{\theta -1} \over (1-u)^\theta} e^{-i uxy} g(\xi) \eqno(23a)$$
with
$$\xi = k (t) y u^\beta (1-u)^{1-\beta} \eqno(23b)$$
for arbitrary functions $k(t)$ of $t$ and $g(\xi)$ of $\xi$ is a solution 
of
$$H  \phi_1 = \lambda_1 \phi_1 \quad , \quad (\theta = {1 \over 2} + i 
\lambda_1) \quad . \eqno(24)$$
Moreover, if $g(0) = 1$ (for example $g(\xi) = \exp (-\xi))$, then ZZFC will 
give the desired boundary condition
$$\phi_1 (x,0) = 0 \eqno(25)$$
for $1 \geq \beta \geq 0$, since $y=0$ implies $\xi =0$. Therefore, with 
the choice of $\theta =z$ and hence $\lambda_1 = \lambda$, the essential 
conditions Eqs. (10) will be obeyed for $\phi = \phi_1$. However, a 
difficulty is that it appears to lead to $<\phi_1 |\phi_1> = \infty$ in 
general, although a possibility may exist to avoid the dilemma for a 
suitable choice of $g(\xi)$. Instead of Eqs. (23), we may also use (by 
letting $u \rightarrow -u$)
$$\eqalignno{\phi_1 (x,y) &= \int^\infty_0 dt\  {t^{z-1} \over 1 + \exp t} 
\int^\infty_0 du\  {u^{\theta -1} \over (1+u)^\theta} e^{iuxy} g(\xi)\quad ,
&(23a^\prime)\cr
\noalign{\vskip 4pt}%
\xi &= k(t) y u^\beta (1+ u)^{1-\beta} \quad , &(23b^\prime)\cr}$$
which satisfies Eq. (24) again. Eq. (25) can also be satisfied although the 
$u$-integration may logarthmically diverge at $u = \infty$ for $y=0$. 
Moreover $<\phi_1 | \phi_1>$ could be even finite,
 at least if Re $\theta > {1 \over 
2}$ for some  $g(\xi)$.  However it seems to be rather unlikely 
that the present $\phi_1$ can offer the correct wave function of the 
problem by the following reason:
The solutions of Eqs. (23) belong to infinite-dimensional realizations of 
SO(1,1). Therefore, the given eigen-value $\lambda$ would then be 
infinitely degenerate because of the Lorentz covariance. We do not know how 
to resolve the dilemma. A simple way is to break the Lorentz invariance of 
H by adding a real non-covariant potential such as $\epsilon y$ for a 
constant $\epsilon$ or by letting $y \rightarrow  y + \epsilon$ for Eq. (8). 
Nevertheless, there exists a indication that the present Hamiltonian H may 
not be completely irrelevant to RH as will be explained below.

The function $f_0(x,y)$ introduced by Eq. (18a)  may also  be related 
to RH by the following reason. We will first state without proof that there 
exist some constants $C_0,\ C_1,\ C_2,$ and $C_3$ such that we have
$$\eqalignno{|f_0 (x,y)| &\leq C_0  &(26a)\cr
 |f_0 (x,y)| &\leq C_1 y^{- {1 \over \beta}\  {\rm Re}\ z} &(26b)\cr
|f_0 (x,y)| &\leq C_2 |x|^{- {1 \over 1-\beta} \ {\rm Re}\ z} &(26c)\cr
|f_0 (x,y)| &\leq C_3 |xy|^{-{\rm Re}\ z} &(26d)\cr}$$
under the assumption of
$$1 > \beta > 0 \quad . \eqno(27)$$
Especially, if we have Re $z> {1 \over 2}$, then $<f_0 |f_0>$ is finite and 
the function $f_0 (x,y)$ will furnish a infinite-dimensional
 \underbar{unitary}
 realization of the Lorentz group SO(1,1) with or without ZZFC. 
Moreover, if ZZFC is assumed, we will have first the orthogonality relation
$$\int^\infty_0 dx \int^\infty_0 dy\  \overline G (xy) f_0 (x,y) = 
\int^\infty_{- \infty} dx \int^\infty_0 dy\  \overline G (xy) f_0 (x,y) = 0
  \eqno(28)$$
for arbitrary function $G (\xi)$ with $\xi = xy$, which will vanish 
sufficiently fast for $\xi \rightarrow \infty$. Second also under ZZFC, it 
satisfies a relation
$$\int^\infty_0 {dk \over k} f_0 \big({x \over k}, ky \big) = 0 \quad . \eqno(29)$$
We can show first Eq. (29) as follows. We rewrite the left side integral of 
Eq. (29) as
$$J = \int^\infty_0 {dk \over k} f_0 \big({x \over k}, ky\big)
 = \int^\infty_0 dt \ 
 t^{z-1} \int^\infty_0 {dk \over k} {\exp (i {x \over k} t^{1-\beta}) \over 1 
+ \exp [t+ky t^\beta]}$$
and change the variable $k$ into $k \rightarrow k^\prime = k t^{\beta -1}$ 
to find
$$J = \int^\infty_0 dt\  t^{z-1} \int^\infty_0 {dk^\prime \over k^\prime} {
\exp (i x/k^\prime) \over 1 + \exp [(1+ k^\prime y) t]} \quad .$$
Interchanging the order of the integral and letting $t \rightarrow t^\prime 
= (1 + k^\prime y)t$, this leads to
$$J = \int^\infty_0 {dk^\prime \over k^\prime} {\exp (ix/k^\prime) \over (1+ 
k^\prime y)^z} \int^\infty_0 dt^\prime {(t^\prime)^{z-1} \over 1 + \exp 
t^\prime}$$
which vanishes identically by ZZFC.

Eq. (28) can then be shown by changing the variable $y$ into $k$ and then 
letting $x \rightarrow \xi = kx$ to calculate
$$\eqalign{\int^\infty_0 dk\  \int^\infty_0 dx \overline G (kx) f_0 (x,k) 
&= \int^\infty_0 dk \int^\infty_0 {d \xi \over k} \overline G (\xi) f_0 
\big( {\xi \over k} , k \big)\cr
&= \int^\infty_0 d \xi\  \overline G (\xi) \int^\infty_0 {dk \over k} f_0 
\big( {\xi \over k} ,k\big)\cr}$$
which is zero by Eq. (29).

The condition Eq. (29) can be interpreted to imply that the 
infinite-dimensional representation space of SO(1,1), spanned by $f_0 (x,y)
$ does \underbar{not} contain any singlet representation of the group.  This 
is because the left side of Eq. (29) is precisely the Lorentz-invariant 
component contained in the representation space, since it is invariant 
under $x \rightarrow {1 \over \alpha} x$ and $y \rightarrow \alpha y$ for 
any positive constant $\alpha$. Then, the orthogonality relation Eq. (28) 
can be readily recognized to be the one between two mutually inequivalent 
representations of SO(1,1) since $G (xy)$ is clearly a Lorentz-scalar. 
Such a relationship between the condition $\zeta (z) =0$ and the absence of 
a trivial representation of SO(1,1) in $f_0 (x,y)$ is quite intriguing and 
may indicate that $f_0 (x,y)$ somehow plays a role in RH.

We can find another type of orthogonality relation between two inequivalent 
eigenfunctions of H. Let $\phi = \phi_1$ and $\psi = f_0$ in Eqs. (9) and 
integrate on $x$ and $y$. If we note $\phi_1 (x,0) = {\partial \over 
\partial x} \phi_1 (x,0) =0$ at $y=0$, together with $H \phi_1 = \lambda_1 
\phi_1$ and $H f_0 = \lambda f_0$, it will lead to
$$\int^\infty_{- \infty} dx \int^\infty_0 dy\  \overline \phi_1 (x,y) f_0 (x,
y) = 0 \eqno(30)$$
provided that we have $\overline \lambda_1 \not= \lambda$.

In order to emphasize the dependence of $f_0 (x,y)$ upon parameters $\beta$ 
and $z$, we now explicitly write it as 
$$F_0 (x,y,z;\beta) = {1 \over \Gamma (z)} \int^\infty_0 dt {t^{z-1} \over
1 + \exp (t+ y t^\beta)} \exp (ix t^{1-\beta}) \eqno(31)$$
so that $f_0 = \Gamma (z) F_0$ and it satisfies the differential equation
$$\bigg\{ {\partial^2 \over \partial x \partial y} + i \beta y {\partial \over
 \partial y} + i ( 1-\beta) x {\partial \over \partial x}\bigg\} F_0 = -i
z F_0  \eqno(32a)$$
as well as
$${\partial \over \partial x} F_0 (x,y,z;\beta) = i
F_0 (x,y,z + 1 -\beta; \beta)\quad . \eqno(32b)$$
For special cases of $\beta=0$ and $\beta=1$, it reproduces the zeta 
function and its generalizations. For $\beta=1$, we change the integration 
variable $t$ into $t^\prime = (1 + y)t$ and note Eq. (1) to obtain
$$F_0 (x,y,z;1) = (1 - 2^{1-z}) \zeta (z) {e^{ix}\over (1 + y)^z} \quad . 
\eqno(33)$$
For $\beta =0$, we calculate
$$F_0 (x,y,z;0) = e^{-y} \Phi (-e^{-y}, z, 1-ix) \eqno(34)$$
where $\Phi (\xi, z, \eta)$ is the generalized zeta function defined by
$$\Phi (\xi, z, \eta) = \sum^\infty_{n=0} (\eta + n)^{-z} \xi^n \eqno(35)$$
which converges for $|\xi| < 1$, $\eta \not= 0, -1, -2, -3, \dots$
When we use the integral representation [8] of
$$\Phi (\xi, z, \eta) = {1 \over \Gamma (z)} \int^\infty_0 dt {t^{z-1} \over 
1 - \xi e^{-t}} e^{-\eta t} \eqno(36)$$
for Re $\eta >0$, Re $z > 0$, $|\xi | \leq 1$, $\xi \not= 1$, and compare 
it with Eq. (31), we find Eq. (34). Since $F_0$ satisfies Eq. (32a), we see 
that $\Phi$ must be a solution of the differential equation
$$\bigg\{ \xi {\partial^2 \over \partial \xi \partial \eta} + \eta {\partial 
\over \partial \eta} \bigg\} \Phi ( \xi, z, \eta) = - z \Phi (\xi , z, \eta) 
\eqno(37)$$
which appears to have been overlooked in literature.

Eq. (37) enjoys symmetries larger than that for Eq. (11) for $\beta \not= 0
$. It is first invariant under
\itemitem{(i)}
$\xi \rightarrow b \xi^k \quad , \quad \eta \rightarrow
{1 \over k} \eta \hfill (38a)$

\noindent for any non-zero constants $k$ and $b$. Second, it also remains invariant 
under a transformation
\itemitem{(ii)} $\Phi (\xi , z, \eta) \rightarrow \hat \Phi (\xi , z,
\eta ) = \xi^\theta \Phi (\xi, z , \eta + \theta) \hfill (38b)$

\noindent for another constant $\theta$. The case of $b=1$ in Eq.
 (38a) reflects the 
original Lorentz invariance Eq. (12). In this connection, identities
$$\eqalignno{\Phi (\xi, z, \eta) + \Phi (-\xi, z, \eta) &= 2^{1-z}
\Phi \big( \xi^2, z, {1 \over 2} \eta \big) \quad , &(39a)\cr
\Phi (\xi, z, \eta) - \Phi (-\xi, z, \eta) &= 2^{1-z}
\xi \Phi \big( \xi^2, z, {1 \over 2}
\big( \eta + 1  \big)\big)  \quad , &(39b)\cr}$$
which can easily be verified from Eq. (35) are clearly consistent with the
 invariances under Eqs. (38)
for special choices of $k=2$, $b=\pm 1$ and $\theta = {1 \over 2}$.

In ending this note, we remark that a special case of $\beta = {1 \over 2}$ 
may be of some interest. In that case, Eq. (8) becomes
$$H = {\partial^2 \over \partial x \partial y} + {i \over 2} \bigg(x {\partial
 \over \partial x} + y {\partial \over \partial y} \bigg) + {i \over 2}
\eqno(40)$$
which is now symmetric in $x$ and $y$. Applying a unitary transformation $H 
\rightarrow \widetilde H$ by
$$\widetilde H = \exp \big({i \over 2} xy\big) H \exp \big(- {i \over 2} xy\big)
 \quad , \eqno(41)$$
it is easy to find
$$\widetilde H = {\partial^2 \over \partial x \partial y} + {1 \over 4} xy 
\quad . \eqno(42)$$
Note first that the constant term ${i \over 2}$ in H has disappeared from 
Eq. (42). Second, $\widetilde H$ is real rather than complex, although this 
property is a special consequence only for $\beta ={1 \over 2}$. Moreover, 
if we change the variables from $x$ and $y$ to $u$ and $v$ given by Eq.
 (13), i.e. $x = u-v$ and $y=u+v$, we can rewrite Eq. (42) now as
$$\widetilde H = {1 \over 4} \bigg({\partial^2 \over \partial u^2} -
 {\partial^2 \over \partial v^2} + u^2 - v^2\bigg) \eqno(43)$$
which represents two-dimensional anti-harmonic oscillator Hamiltonian in 
the sense that the signs of the quadratic potentials have the wrong signs 
in comparison to the harmonic oscillator case.

We have also found the following rather peculiar solution of $H \phi = 
\lambda \phi$ for the case of $\beta = {1 \over 2}$. The function
$$\phi (x,y) = {\xi^{2 z} \over \xi^2 +1} \exp \bigg(- {i \over 2} x \xi
\bigg) \eqno
(44a)$$
with
$$\xi = y + (1 + y^2)^{{1 \over 2}} \eqno(44b)$$
can be shown to satisfy $H \phi = \lambda \phi$, although  it may have 
nothing to do with the problem of RH.

In conclusion, we have attempted in this note to present some arguments for 
possible relevance of our Hamiltonians to RH. Although they may not be the 
ultimate answer to the problem, there are at least some indications that 
they may be indirectly useful.

\noindent \underbar{\bf Acknowledgement}

This paper is supported in part by the U.S. Department of Energy Grant no. 
DE-~FG02-91ER40685.

\vfill\eject

\noindent \underbar{\bf References}

\item{1.} Titschmarsh, E.C.; 1951 \lq\lq The Theory of the Riemann Zeta 
Function", (London, Oxford Unversity Press).

\item{2.} Edwards, H.M.; 1974 \lq\lq Riemann Zeta Function", (New York, 
Academic Press).

\item{3.} Karatsuba, A.A.; 1995 \lq\lq Complex Analysis in Number Theory", 
(Boca Raton, Florida, CRC Press).

\item{4.} Odlyzko, A.M.; 1987, Math. of Comp. {\bf 48}, 273.

\item{5.} Mehta, M.L.; 1990 \lq\lq Random Matrices and the Statistical 
Theory of Energy Level", revised and enlarged Second Edition, (New York, 
Academic Press).

\item{6.} Gutzwiller, M.C.; 1990 \lq\lq Chaos in Classical and Quantum 
Physics", (New York, Springer Verlag).

\item{7.} Bolte, J.; 1993, Int. J. Mod. Phys. {\bf B7}, 4451.

\item{8.} Erd\'elyi, A., Magnus, W., Oberhettinger, F., and Tricomi, F.G.; 
1953 \lq\lq Higher Transcendental Functions, Vol.~I Bateman Manuscript 
Project", (New York, McGraw Hill).

\end